\documentclass[fleqn,12pt,twoside]{article} 
\usepackage{espcrc1} 
 
\usepackage{graphicx} 
 
\newcommand{\AmS}{{\protect\the\textfont2 
  A\kern-.1667em\lower.5ex\hbox{M}\kern-.125emS}} 
 
\newcommand{\lsim}{\mathrel{\rlap{\lower4pt\hbox{\hskip0pt$\sim$}} 
\raise1pt\hbox{$<$}}}           
\newcommand{\gsim}{\mathrel{\rlap{\lower4pt\hbox{\hskip0pt$\sim$}} 
\raise1pt\hbox{$>$}}}           
 
\newcommand{\sfrac}[2]{\mbox{\footnotesize $\frac{#1}{#2}$}} 

\title{Dyson-Schwinger Equations: An Instrument for Hadron Physics%
\thanks{CDR thanks the organisers for their hospitality and support.
This work was also supported by: 
{\it FWF Erwin-Schr\"odinger-Auslandsstipendium Nr.}\ J2233-N08;
Department of Energy Nuclear Physics Division contract no.\ W-31-109-ENG-38;
and the A.v.\ Humboldt Foundation via a F.W.\ Bessel Research Award.}
} 
 
\author{A.\ Krassnigg\,\address[ANL]{%
Physics Division, Argonne National Laboratory, Argonne IL 60439, 
USA}%
\ and C.\,D.\ Roberts\,\addressmark[ANL]$^{,}$\address{%
Fachbereich Physik, Universit\"at Rostock, D-18051 Rostock, Germany}} 
 
\begin{document} 
 
\maketitle 
 
\begin{abstract} 
Dyson-Schwinger equations furnish a Poincar\'e covariant approach to hadron
physics. 
They reveal that dynamical chiral symmetry breaking is tied to the long-range
behaviour of the strong interaction and make predictions corroborated by
modern lattice-QCD simulations.  A hallmark in the contemporary use of DSEs
is the existence of a nonperturbative, symmetry preserving truncation that
enables the proof of exact results.  The systematic error associated with the
truncation's leading term has been quantified and this underpins an
efficacious one-parameter model that is being employed to study meson excited
states.
\end{abstract} 
 
\section{DRESSED-QUARK PROPAGATOR} 
\setcounter{footnote}{0} 
The Dyson-Schwinger equations (DSEs) are a nonperturbative approach to
studying continuum QCD \cite{revbastirevreinhard,revpieter}.  At the simplest
level they provide a generating tool for perturbation theory and, because QCD
is asymptotically free, this feature has the potential to materially reduce
any model dependence in DSE applications.  This comment is explained by
noting that the long-range behaviour of the interaction between light-quarks
is hitherto unknown and must therefore be modelled.  Hence, the fact that a
weak coupling expansion of the DSEs reproduces every diagram in perturbation
theory entails that in this approach the model dependence is restricted to
the infrared domain $k^2\lsim 1\,$GeV$^2$.
 
The DSEs are a nonperturbative tool and can therefore be used directly to
study: hadrons as bound states; the importance of dynamical chiral symmetry
breaking (DCSB); and the confinement of quarks and gluons.  These phenomena
are all linked to the infrared behaviour of the interaction.  Hence the DSEs
provide a bridge between this behaviour and observables, and thereby a means
by which it can be charted.
 
The best known DSE is the simplest: the \textit{gap} equation, which
describes how a fermion's propagator is modified by interactions with the
medium being traversed.  In QCD that equation is the DSE for the
dressed-quark propagator:
\begin{equation} 
\label{gendse} S(p)^{-1} = Z_2 \,(i\gamma\cdot p + m_{\rm bare}) +\, Z_1 
\int^\Lambda_q \, g^2 D_{\mu\nu}(p-q) \frac{\lambda^a}{2}\gamma_\mu S(q) 
\Gamma^a_\nu(q;p) \,,
\end{equation} 
wherein: $D_{\mu\nu}(k)$ is the dressed-gluon propagator; $\Gamma^a_\nu(q;p)$
is the dressed-quark-gluon vertex; $m_{\rm bare}$ is the
$\Lambda$-de\-pen\-dent current-quark bare mass; and $\int^\Lambda_q :=
\int^\Lambda d^4 q/(2\pi)^4$ represents a translationally-invariant
regularisation of the integral, with $\Lambda$ the regularisation mass-scale.
In addition, $Z_{1,2}(\zeta^2,\Lambda^2)$ are the quark-gluon-vertex and
quark wave function renormalisation constants, which depend on $\Lambda$ and
the renormalisation point, $\zeta$, as does the mass renormalisation constant
$ Z_m(\zeta^2,\Lambda^2) = Z_4(\zeta^2,\Lambda^2)/Z_2(\zeta^2,\Lambda^2) $.
The solution has the form
\begin{equation} 
 S^{-1} (p) =  i \gamma\cdot p \, A(p^2,\zeta^2) + B(p^2,\zeta^2) 
 \equiv \frac{1}{Z(p^2,\zeta^2)}\left[ i\gamma\cdot p + M(p^2)\right], 
\label{sinvp} 
\end{equation} 
where $M(\zeta^2) \equiv m(\zeta):= m_{\rm 
bare}(\Lambda)\, Z_m^{-1}(\zeta^2,\Lambda^2)$ is the running quark mass. 
 
The gap equation illustrates the features and flaws of each DSE.  It is a
nonlinear integral equation for $S(p)$ and hence can yield much-needed
nonperturbative information.  However, the kernel involves the two-point
function $D_{\mu\nu}(k)$ and the three-point function
$\Gamma^a_\nu(q;p)$. The gap equation is therefore coupled to the DSEs these
functions satisfy. Those equations in turn involve higher $n$-point functions
and hence the DSEs are a tower of coupled integral equations with a tractable
problem obtained only once a truncation scheme is specified.  It is
unsurprising that the best known truncation scheme is the weak coupling
expansion, which reproduces every diagram in perturbation theory.  This
scheme is systematic and valuable in the analysis of large momentum transfer
phenomena because QCD is asymptotically free but it precludes any possibility
of obtaining nonperturbative information, which we identified as a key aspect
of the DSEs.
 
In spite of the problem with truncation, gap equations have long been used 
efficaciously in obtaining nonperturbative information about many-body 
systems.  The positive outcomes have been achieved through the simple 
expedient of employing a rudimentary truncation and comparing the results 
with observations.  Naturally, agreement under these circumstances is not 
compelling evidence that the contributions omitted are small.
However, it does justify further study, and an accumulation of good results
is grounds for a concerted attempt to substantiate a reinterpretation of the
truncation as the first term in a systematic and reliable approximation.
That has recently been achieved \cite{benderdetmold}.
 
\begin{figure}[t]
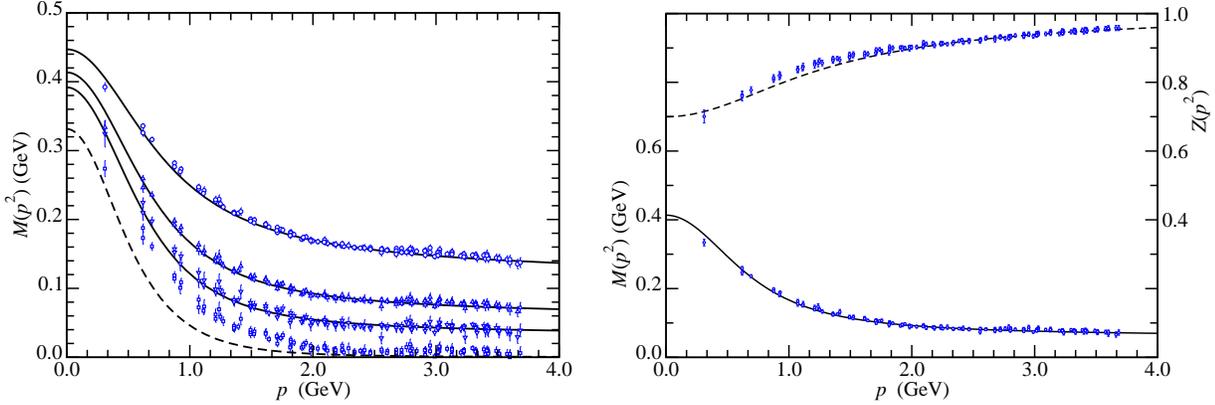
 
\begin{minipage}[t]{0.47\textwidth} 
\includegraphics[height=0.71\textwidth]{fig1.eps} 
\end{minipage} 
\hspace*{0.4em} 
\begin{minipage}[t]{0.47\textwidth} 
\includegraphics[height=0.7\textwidth]{fig2.eps} 
\end{minipage}\vspace{-4ex} 
\caption{\label{mandarcflattice} \textit{Left Panel} -- Dashed-curve: gap
equation's solution in the chiral limit; solid curves: solutions for $M(p^2)$
obtained using $m(\zeta)$ in Eq.\ (\protect\ref{amvalues}). (From Ref.\
\protect\cite{mandar}.)  Data, upper three sets: lattice results for $M(p^2)$
in GeV at $am$ values in Eq.\,(\protect\ref{amvalues}); lower points (boxes):
linear extrapolation of these results \protect\cite{bowman2} to $a m
=0$. \textit{Right Panel} -- Dashed curve, $Z(p^2)$, and solid curve,
$M(p^2)$ calculated from the gap equation with $m(\zeta)=55\,$MeV
\protect\cite{mandar}.  Data, quenched lattice-QCD results for $M(p^2)$ and
$Z(p^2)$ obtained with $am = 0.036$ \protect\cite{bowman2}.}
\end{figure}

The gap equation's solution has long been of interest in grappling with DCSB
and typical results are depicted in Fig.\,\ref{mandarcflattice}.  The
infrared suppression of $Z(p^2)$ and enhancement of $M(p^2)$ evident in this
figure are longstanding predictions of DSE studies \cite{cdragw}, and in this
behaviour one perceives directly the evolution of a current-quark into a
constituent-quark.  A critical feature is that, so long as the kernel of the
gap equation has sufficient integrated strength on the infrared domain, a
nonzero running quark mass is obtained even in the chiral limit.  This effect
is DCSB. It is impossible at any finite order of perturbation theory and also
apparent in Fig.\ \ref{mandarcflattice}.  The precise connection between the
dressed-quark propagator and vacuum quark condensate is discussed in
Ref.\,\cite{kurt}, and a connection with quark confinement is canvassed in
Ref.\,\cite{mandar}.
 
The dressed-quark propagator can be calculated in lattice-regularised QCD. 
Results are available in the quenched truncation, and depicted in Fig.\ 
\ref{mandarcflattice} are those of Ref.\ \cite{bowman2} obtained with the 
current-quark masses ($\zeta = 19\,$GeV) 
\begin{equation} 
\label{amvalues} 
\begin{array}{l|lll} 
a\,m_{\rm lattice} & 0.018 & 0.036 & 0.072 \\\hline m(\zeta) ({\rm GeV}) & 
0.030 & 0.055 & 0.110 
\end{array}\,. 
\end{equation} 
The precise agreement with DSE results is not accidental.  (NB.\ The chiral
limit discrepancy points to the inadequacy of a linear extrapolation of
lattice data to this limit \cite{mandar}.) The essential agreement between
lattice results and DSE predictions was highlighted in
Refs.\,\cite{raya,pctlattice} but Ref.\,\cite{mandar} pursued a different
goal.  Only recently has reliable information about the gap equation's kernel
at infrared momenta begun to emerge, in the continuum \cite{alkofer} and on
the lattice \cite{latticegluon}. Reference \cite{mandar} therefore employed
an \textit{Ansatz} for the infrared behaviour of the gap equation's kernel in
order to demonstrate that it is possible to correlate lattice results for the
gluon and quark Schwinger functions via QCD's gap equation.  This required
the gap equation's kernel to exhibit infrared enhancement over and above that
observed in the gluon propagator alone, which could be attributed to an
amplification of the dressed-quark-gluon vertex whose magnitude is consistent
with that observed in quenched lattice estimates of this three-point function
\cite{ayse}.
 
\section{HADRONS} 
It is evident that reliable knowledge of QCD's two-point functions (the 
propagators for QCD's elementary excitations) is available.  Direct comparison 
with experiment requires an equally good understanding of bound states. 
Progress here has required the evolution of an understanding of the intimate 
connection between symmetries and DSE truncation schemes.  This is well 
illustrated by considering the pion, whose properties are profoundly connected 
with DCSB.  Indeed, the correct understanding of pion observables requires an 
approach to contain a well-defined and valid chiral limit. 
 
Chiral symmetry and its breaking are expressed through the axial-vector 
Ward-Ta\-ka\-ha\-shi identity 
\begin{equation} 
\label{avwtim} P_\mu \Gamma_{5\mu}^j(k;P)  = {\cal S}^{-1}(k_+) i 
\gamma_5\frac{\tau^j}{2} +  i \gamma_5\frac{\tau^j}{2} {\cal S}^{-1}(k_-) - i 
{\cal M}(\zeta) \,\Gamma_5^j(k;P) - \Gamma_5^j(k;P)\,i {\cal M}(\zeta), 
\end{equation} 
where $k_\pm = k \pm P/2$, and $\{\tau^j,j=1,2,3\}$ are the Pauli matrices as 
herein we focus on $SU(2)$-flavour.  This identity connects the axial-vector 
vertex: $\Gamma_{5\mu}^j(k;P)$, $P$ is the total momentum, with the dressed 
quark propagator: ${\cal S} = {\rm diag}[S_u,S_d]$, the pseudoscalar vertex: 
$\Gamma_{5}^j(k;P)$, and the current-quark mass matrix: ${\cal M}(\zeta) = 
{\rm diag}[m_u(\zeta),m_d(\zeta)]$.  The propagator satisfies the gap 
equation but the vertices are determined by inhomogeneous Bethe-Salpeter 
equations; e.g., 
\begin{eqnarray} 
\label{genave} \left[\Gamma_{5\mu}^j(k;P)\right]_{tu} &= &Z_2 \, 
\left[\gamma_5\gamma_\mu \frac{\tau^j}{2}\right]_{tu} \,+ \int^\Lambda_q \, 
[{\cal S}(q_+) \Gamma^j_{5\mu}(q;P) {\cal S}(q_-)]_{sr} 
\,K^{rs}_{tu}(q,k;P)\,, 
\end{eqnarray} 
wherein $K(q,k;P)$ is the dressed-quark-antiquark scattering kernel.  The
importance of DCSB entails that any truncation useful in understanding low
energy phenomena must be nonperturbative and preserve Eq.\,(\ref{avwtim})
without fine tuning. This constraint requires an intimate connection between
$K(q,k;P)$ and the gap equation's kernel.
 
\begin{figure}[t] 
\centerline{\resizebox{0.70\textwidth}{!}{\includegraphics{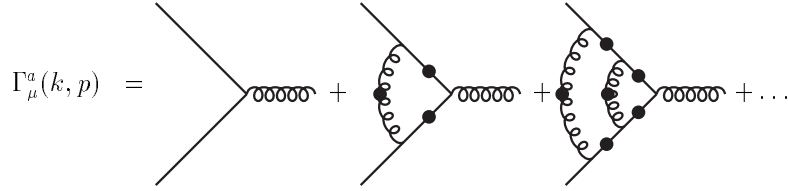}}}\vspace{-4ex} 
 
\caption{\label{fig3} Planar dressed-quark-gluon vertex obtained by neglecting 
contributions associated with explicit gluon self-interactions. Solid circles 
indicate fully dressed propagators.  The quark-gluon vertices are not dressed.} 
\end{figure} 
 
One systematic and nonperturbative truncation scheme has been identified that 
explicates this connection, and hence preserves QCD's global symmetries 
\cite{benderdetmold}.  It is a dressed-loop expansion of the 
dressed-quark-gluon vertices that appear in the half-amputated 
dressed-quark-anti\-quark scattering matrix: $S^2 K$. In this scheme, as in 
perturbation theory, it is impossible, in general, to obtain complete 
closed-form expressions for the kernels of the gap and Bethe-Salpeter 
equations.  However, for the planar dressed-quark-gluon vertex depicted in 
Fig.\ \ref{fig3}, closed forms can be obtained and a number of significant 
features illustrated \cite{benderdetmold} when one uses the following model 
for the dressed-gluon line \cite{mn83} 
\begin{equation} 
\label{mnmodel} g^2 \, D_{\mu\nu}(k) = \left[\delta_{\mu\nu} - \frac{k_\mu 
k_\nu}{k^2}\right] (2\pi)^4\, {\cal G}^2 \, \delta^4(k)\,, 
\end{equation} 
where ${\cal G}$ sets the model's mass-scale. This form has many positive
attributes in common with the class of renormalisation-group-improved models
and furthermore its particular momentum-dependence works to advantage in
reducing integral equations to character-preserving algebraic equations.
 
It is a general result \cite{benderdetmold} that with any vertex whose 
diagrammatic content is explicitly known; e.g., Fig.\,\ref{fig3}, it is 
always possible to construct a unique Bethe-Salpeter kernel which ensures the 
Ward-Takahashi identities are automatically fulfilled.  That kernel is 
necessarily non-planar.  This becomes transparent with the model in 
Eq.\,(\ref{mnmodel}), using which the gap equation obtained with the vertex 
depicted in Fig.\,\ref{fig3} reduces to an algebraic equation, irrespective 
of the number of dressed-gluon rungs that are retained, and the same is true 
of the Bethe-Salpeter equations in every channel: pseudoscalar, vector, etc. 
 
\begin{table}[t] 
\caption{\label{tab1} $\pi$ and $\rho$ meson masses obtained with ${\cal G}= 
0.48\,{\rm GeV}$.   (Dimensioned quantities in GeV.)  $n$ is the number of 
dressed-gluon rungs retained in the planar vertex, see Fig.~\protect\ref{fig3}, 
and hence the order of the vertex-consistent Bethe-Salpeter kernel.} 
\begin{tabular*} 
{\hsize}{l@{\extracolsep{0ptplus1fil}} 
|c@{\extracolsep{0ptplus1fil}}c@{\extracolsep{0ptplus1fil}} 
c@{\extracolsep{0ptplus1fil}}c@{\extracolsep{0ptplus1fil}}} 
\hline
%
 & $M_H^{n=0}$ & $M_H^{n=1}$ & $M_H^{n=2}$ & $M_H^{n=\infty}$\\\hline 
$\pi$, $m=0$ & 0 & 0 & 0 & 0\\ 
$\pi$, $m=0.011$ & 0.152 & 0.152 & 0.152 & 0.152\\\hline 
$\rho$, $m=0$ & 0.678 & 0.745 & 0.754 & 0.754\\ 
$\rho$, $m=0.011$ & 0.695 & 0.762 & 0.770 & 0.770 \\ 
%
\hline 
\end{tabular*} 
\end{table} 
 
Results for the $\pi$ and $\rho$ are illustrated in Table \ref{tab1}.  It is
evident that, irrespective of the order of the truncation; i.e., the number
of dressed gluon rungs in the quark-gluon vertex, the pion is massless in the
chiral limit.  (NB.\ This pion is composed of heavy dressed-quarks, as is
evident in the calculated scale of the dynamically generated dressed-quark
mass function: $M(0) \approx 0.5\,$GeV; viz., dynamically generated
constituent-quarks compose the pion.) The masslessness of the $\pi$ is a
model-independent consequence of the consistency between the Bethe-Salpeter
kernel and the kernel in the gap equation.  Furthermore, the bulk of the
$\rho$-$\pi$ mass splitting is present in the chiral limit and with the
simplest ($n=0$; i.e., rainbow-ladder) kernel, which makes plain that this
mass difference is driven by the DCSB mechanism: it is not the result of a
finely adjusted hyperfine interaction.  Finally, the quantitative effect of
improving on the rainbow-ladder truncation; viz., including more
dressed-gluon rungs in the gap equation's kernel and consistently improving
the kernel in the Bethe-Salpeter kernel, is a 10\% correction to the vector
meson mass.  Simply including the first correction ($n=1$; i.e., retaining
the first two diagrams in Fig.\ \ref{fig3}) yields a vector meson mass that
differs from the fully resummed result by
$\stackrel{<}{\mbox{\tiny$\sim$}}1$\%.  The rainbow-ladder truncation is
clearly accurate in these channels.
 
\section{{\it AB INITIO} CALCULATIONS} 
The leading-order term in the truncation scheme described above is the 
renormalisation-group-improved rainbow-ladder truncation, which as we have 
just seen is reliable for flavour nonsinglet pseudoscalar and vector mesons. 
This truncation has been refined and exploited by Maris and Tandy in a series 
of articles\,\footnote{An exemplary success was their prediction 
\protect\cite{maristandypion} of the electromagnetic pion form factor 
\protect\cite{volmer}.  In addition, see Ref.\,\protect\cite{maristandy} and 
references therein, and the review of their substantial contributions in 
Ref.\,\protect\cite{revpieter}.}  via a one-parameter model for the behaviour 
of the dressed-quark-quark interaction: 
\begin{equation} 
\label{alphamt} 
\frac{\alpha(Q^2)}{Q^2} = \frac{\pi}{\omega^6} D\, Q^2 {\rm e}^{-Q^2/\omega^2} 
+ \,\frac{ \pi\, \gamma_m } { \frac{1}{2}\ln\left[\tau + \left(1 + 
Q^2/\Lambda_{\rm QCD}^2\right)^2\right]} \, {\cal F}(Q^2) \,, \label{gk2} 
\end{equation} 
wherein ${\cal F}(Q^2)= [1 - \exp(-Q^2/[4 m_t^2])]/Q^2$, $m_t$ $=$ 
$0.5\,$GeV; $\tau={\rm e}^2-1$; $\gamma_m = 12/25$; and $\Lambda_{\rm QCD} 
=0.234\,$GeV.  This simple form expresses the interaction as a sum: the 
second term ensures that perturbative behaviour is preserved at short-range; 
and the first makes provision for enhancement at long-range.  The true 
parameters in Eq.\,(\ref{gk2}) are $D$ and $\omega$, which together determine 
the integrated infrared strength of the rainbow-ladder kernel.  However, they 
are not independent: in fitting to a selection of observables, a change in 
one is compensated by altering the other; e.g., on the domain 
$\omega\in[0.3,0.5]\,$GeV, the fitted observables are approximately constant 
along the trajectory \cite{raya} $\omega \,D = (0.72\,{\rm GeV})^3$. Hence 
Eq.\,(\ref{gk2}) is a one-parameter model.  This correlation: a reduction in 
$D$ compensating an increase in $\omega$, ensures a fixed value of the 
integrated infrared strength.  It also defines a single dressed-glue 
mass-scale, $m_g = 720\,$MeV, which characterises infrared gluodynamics. 
 
A new application is the calculation of properties of meson excited states, 
which is a first step in seeking exotics and hybrids.  As an example, it is a 
general feature of QCD that the axial-vector vertex, obtained via 
Eq.\,(\ref{genave}), exhibits a pole whenever $P^2= - m^2_{\pi_n}$, where 
$m_{\pi_n}$ is the mass of any pseudoscalar $u$-$d$ meson; viz., 
\begin{equation} 
\left. \Gamma_{5 \mu}^j(k;P)\right|_{P^2+m_{\pi_n}^2 \approx 0} = 
\mbox{regular\ terms} +  \frac{f_{\pi_n} \, P_\mu}{P^2 + 
m_{\pi_n}^2} \Gamma_{\pi_n}^j(k;P)\,, \label{genavv} 
\end{equation} 
with $\Gamma_{\pi_n}^j(k;P)$ being the $0^{-+}$ bound state's Bethe-Salpeter 
amplitude: 
\begin{eqnarray} 
\nonumber \Gamma_{\pi_n}^j(k;P)& =&  \tau^j \gamma_5 \left[ i E_{\pi_n}(k;P) 
+ \gamma\cdot P F_{\pi_n}(k;P) \right. \\ 
&&  \left.  +\,  \gamma\cdot k \,k \cdot P\, G_{\pi_n}(k;P) + 
\sigma_{\mu\nu}\,k_\mu P_\nu \,H_{\pi_n}(k;P)  \right], \label{genpibsa} 
\end{eqnarray} 
and $f_{\pi_n}$, its leptonic decay constant
\begin{equation} 
\label{fpin} f_{\pi_n} \,\delta^{ij} \,  P_\mu = Z_2\,{\rm tr} \int^\Lambda_q 
\sfrac{1}{2} \tau^i \gamma_5\gamma_\mu\, {\cal S}(q_+) \Gamma^j_{\pi_n}(q;P) 
{\cal S}(q_-) \,, 
\end{equation} 
where the trace is over colour, flavour and spinor indices.  The lowest mass 
pole contribution, denoted by $n=0$, is the ground state pion, which receives 
the spectroscopic assignment $N \; ^{2 s +1}\!L_J = 1 \, ^1\!S_0$ in the 
naive $q \bar q$ quark model.  The $n\geq 1$ pseudoscalar meson poles 
correspond to those excited states of the pion that would receive the 
spectroscopic assignments $(n+1)\, ^1\!S_0$ in the quark model. 
 
The pseudoscalar vertex, $\Gamma_5^j(k;P)$, which appears in 
Eq.\,(\ref{avwtim}), also exhibits such a pole: 
\begin{equation} 
 \left. i \Gamma_{5}^j(k;P)\right|_{P^2+m_{\pi_n}^2 \approx 0} = 
\mbox{regular\ terms} + \frac{ \rho_{\pi_n} }{P^2 + m_{\pi_n}^2}\, 
\Gamma_{\pi_n}^j(k;P)\,,\label{genpvv} 
\end{equation} 
\begin{equation} 
\label{cpres} i  \rho_{\pi_n}\!(\zeta)\, \delta^{ij}  = Z_4\,{\rm tr} 
\int^\Lambda_q \sfrac{1}{2} \tau^i \gamma_5 \, {\cal S}(q_+) 
\Gamma^j_{\pi_n}(q;P) {\cal S}(q_-)\,. 
\end{equation} 
 
In QCD it therefore follows as a necessary consequence of chiral symmetry and 
its dynamical breaking that for any pseudoscalar $u$-$d$ meson 
\cite{krassnigg} 
\begin{equation} 
\label{gmorgen} f_{\pi_n} m_{\pi_n}^2 = [ m_u(\zeta) + m_d(\zeta) ] \, 
\rho_{\pi_n}(\zeta)\,. 
\end{equation} 
(The generalisation to $SU(N_f)$-flavour is straightforward, following 
Ref.\,\cite{mr97}.)  The so-called Gell-Mann--Oakes--Renner relation for the 
ground state pion appears as a corollary of Eq.\,(\ref{gmorgen}) 
\cite{mrt98}, and another important corollary of Eq.\,(\ref{gmorgen}), valid 
for pseudoscalar mesons containing at least one heavy-quark, is described in 
Ref.\,\cite{hqlimit}.  In addition, the proof of Eq.\,(\ref{gmorgen}) 
establishes \cite{krassnigg} that in the chiral limit 
\begin{equation} 
\label{fpiniszero} 
f^0_{\pi_{n\neq 0}}:= \lim_{m \to 0} \, f_{\pi_{n\neq 0}} = 0 \,; 
\end{equation} 
i.e., in the chiral limit the leptonic decay constant vanishes for every 
pseudoscalar meson except the pion.  This exact result is a constraint on 
models and methods used to search for exotics and hybrids. 
 
The rainbow-ladder truncation obtained with the interaction of 
Eq.\,(\ref{alphamt}) provides an excellent description of ground state 
pseudoscalar mesons \cite{revpieter}.  Nevertheless, the long-range part of 
this interaction expresses a model and its veracity can be tested further by 
studying excited pseudoscalar mesons.  We have begun this process, focusing 
on the $n=1$ pseudoscalar meson ($2\, ^1\!S_0$ in the quark model). 
 
In our studies to date we have only retained the $i\gamma_5 E_{\pi_1}(k;P)$ 
piece in Eq.\,(\ref{genpibsa}).  Employing this expedient for the ground 
state pion leads to a $25$\% underestimate of $m_{\pi_0}$.  Moreover, in 
neglecting the pseudovector components, we omit those terms which in the 
pion's rest frame signal the presence of quark orbital angular momentum.  In 
QCD they are necessarily nonzero \cite{mrt98,mrpion}. 
 
To determine $m_{\pi_{n=1}}$ we first solved a rainbow truncation of the gap 
equation; viz., 
\begin{equation} 
S(p)^{-1} = Z_2 \,(i\gamma\cdot p + m_{\rm bare}) +\, \int^\Lambda_q \, 
4\pi\alpha((p-q)^2)\,D^{\rm free}_{\mu\nu}(p-q) \frac{\lambda^a}{2}\gamma_\mu 
S(q) \frac{\lambda^a}{2}\gamma_\nu \,, 
\end{equation} 
where $D^{\rm free}_{\mu\nu}(k)$ is the undressed gauge boson propagator. 
This allows one to complete the specification of the ladder-truncation 
Bethe-Salpeter equation's kernel: 
\begin{equation} 
\label{bsemod} 
\Gamma_{\pi_{n}}(k;P) + \int^\Lambda_q 4\pi\,\alpha((k-q)^2)\, 
D_{\mu\nu}^{\rm free}(k-q) \frac{\lambda^a}{2}\gamma_\mu\, {\cal 
S}(q_+)\,\Gamma_{\pi_{n}}(q;P)\,{\cal S}(q_-)\, \frac{\lambda^a}{2}\gamma_\nu 
= 0\,, 
\end{equation} 
which can be solved as described in Ref.\,\cite{krassnigg} to obtain the mass 
of the ground state and excited pseudoscalar mesons. 
 
\begin{figure}[t] 
\centerline{\resizebox{0.70\textwidth}{!}{\includegraphics[angle=-90]{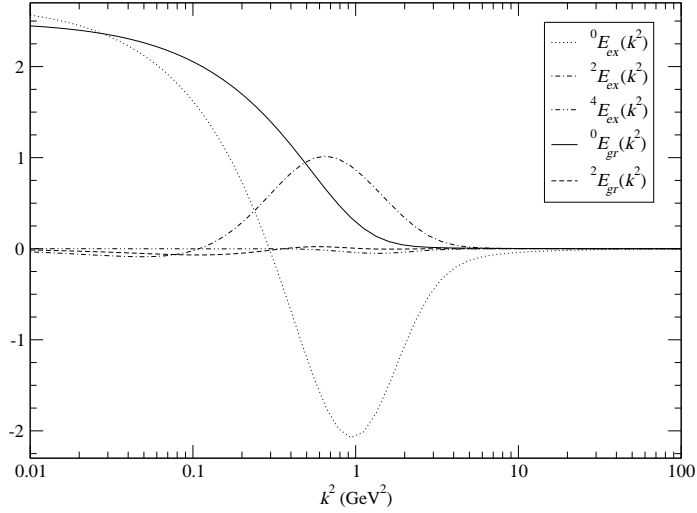}}}\vspace{-4ex} 
 
\caption{\label{fig4} Chebyshev moments, Eq.\,(\ref{chebexp}), of
$E_{\pi_0}(k;P)\equiv E_{gr}(k;P)$ and $E_{\pi_1}(k;P)\equiv
E_{ex}(k;P)$. The leading moment is sufficient to accurately represent the
ground state while a veracious description of the first excited state
requires inclusion of the second moment.  Note the appearance of a single
node in each of the first excited state's Chebyshev moments.  (The amplitudes
are canonically normalised.)}
\end{figure} 
 
One should note that in relativistic quantum field theory the scalar 
functions in Eq.\,(\ref{genpibsa}) depend on two variables: $k^2$, $k\cdot 
P$.  In quantum mechanics $k\cdot P \equiv 0$, of course, because of 
``on-shell'' constraints on the constituents, and this degree of freedom is 
absent.  As a matter of practice, we solve Eq.\,(\ref{bsemod}) by using a 
Chebyshev decomposition; i.e., we write\,\footnote{The Bethe-Salpeter 
amplitude for charge conjugation eigenstates is even under $k\cdot P \to 
-k\cdot P$ and hence only even Chebyshev moments contribute.} 
\begin{equation} 
\label{chebexp} 
E_{\pi_n}(k^2,k\cdot P;P^2) =  \sum_{i=0,2,4,\ldots}^{N_{\rm 
 max}} \,^i\!E_{\pi_n}(k^2;P^2)\,U_i(\cos\beta)\,, 
\end{equation} 
where $\{U_i(x);i=0,\ldots,\infty\}$ are Chebyshev polynomials of the second 
kind, and increase $N_{\rm max}$ until $m_{\pi_n}$ stabilises.  This 
procedure has the merit of limiting the amount of computer memory required. 
It also converges very rapidly for the ground state pion, with only the 
leading Chebyshev moment being required \cite{mr97}, as apparent in 
Fig.\,\ref{fig4}. 
 
We have obtained the mass and amplitude for the ground state pion, using the 
complete expression in Eq.\,(\ref{genpibsa}), and found: $ f_{\pi_0} = 
0.092$\,{\rm GeV}; $m_{\pi_0} = 0.14$\,{\rm GeV}; $\rho_{\pi_0} = (0.81\,{\rm 
GeV})^2$, at a current-quark mass $m_d(1\,{\rm GeV})= m_u(1\,{\rm GeV})= 
5.5\,$MeV, reproducing the results in Ref.\,\cite{maristandypion}.  We can 
report in addition that using solely the $i\gamma_5 E_{\pi_1}(k;P)$ term in 
Eq.\,(\ref{genpibsa}) 
\begin{equation} 
m_{\pi_1} = 1.1\,{\rm GeV}\,, 
\end{equation} 
cf.\ $m_{\pi_1}^{\rm expt.} = 1.3\pm 0.1 \,{\rm GeV}$.  In omitting the
pseudoscalar meson's pseudovector components one violates Eq.\,(\ref{avwtim})
\cite{mr97} and hence we are not yet in a position to verify that
Eq.\,(\ref{fpiniszero}) is preserved in the rainbow-ladder truncation.
Nevertheless, we estimate $f_{\pi_{n=1}} \leq 1.5\,{\rm MeV}$ at the physical
current-quark mass \cite{krassnigg}.
 
We plot the leading Chebyshev moments of $E_{\pi_1}(k;P)$ in 
Fig.\,\ref{fig4}, wherefrom it appears that the first two moments are 
sufficient for an accurate representation of this amplitude but confirmation 
must await the inclusion of the pseudovector components.  It is worth 
remarking that the moments of $E_{\pi_1}(k;P)$ each possess a single node and 
exhibit tails that extend to larger values of $k^2$ than their analogue in 
the ground state.  These observations are redolent of those one could make in 
comparing Fourier transforms of the radial wave functions for bound states in 
a potential well.  This fact emphasises the intuitive character of 
Bethe-Salpeter amplitudes. 
 
\section{EPILOGUE} 
There are many additional applications of interest to this community, among 
them: the \textit{ab initio} calculation of electroweak and transition form 
factors for other mesons, and $\pi \pi$ scattering \cite{pipi}; and a 
calculation of the pion's valence-quark distribution function, whose 
discrepancy with extant data raises difficult questions \cite{cdrpiv}. 
%
A pressing contemporary challenge is the extension of the framework to the 
calculation of baryon observables, aspects of which are beginning to be 
understood \cite{piN}.

\end{document}